\documentclass[unpublished,a4paper,onecolumn,superscriptaddress,11pt]{quantumarticle}
\pdfoutput=1
\usepackage{amsmath}
\usepackage{latexsym}
\usepackage{amssymb}
\usepackage{graphicx,epstopdf,subfigure,color,hyperref,enumerate}
\usepackage[normalem]{ulem}
\usepackage{braket}
\usepackage{float}
\usepackage{caption}
\usepackage{multirow}

\definecolor{Red}{rgb}{1,0,0}

\begin{document}
\title{Interpreting weak value amplification with a toy realist model}
\author{Josiah Sinclair }
\affiliation{Centre for Quantum Information and Quantum Control and Institute for Optical Sciences,
Department of Physics, University of Toronto, 60 St. George Street, Toronto, Ontario M5S 1A7, Canada.}
\orcid{..}
\email{sinclairjosiah@gmail.com}
\author{David Spierings}
\affiliation{Centre for Quantum Information and Quantum Control and Institute for Optical Sciences,
Department of Physics, University of Toronto, 60 St. George Street, Toronto, Ontario M5S 1A7, Canada.}
\author{Aharon Brodutch}
\affiliation{Centre for Quantum Information and Quantum Control and Institute for Optical Sciences,
Department of Physics, University of Toronto, 60 St. George Street, Toronto, Ontario M5S 1A7, Canada.}
\affiliation{The Edward S. Rogers Sr. Department of
Electrical \& Computer Engineering
10 King's College Road
Toronto, Ontario  M5S 3G4, Canada.}
\author{Aephraim M. Steinberg}
\affiliation{Centre for Quantum Information and Quantum Control and Institute for Optical Sciences,
Department of Physics, University of Toronto, 60 St. George Street, Toronto, Ontario M5S 1A7, Canada.}
\affiliation{Canadian Institute For Advanced Research, MaRS Centre,
West Tower 661 University Ave., Toronto, Ontario M5G 1M1, Canada.}

\begin{abstract}

Constructing an ontology for quantum theory is challenging, in part due to unavoidable measurement back-action. The Aharonov-Albert-Vaidman weak measurement formalism provides a method to predict  measurement results (weak values) in a regime where back-action is negligible. The weak value appears analogous to a classical conditional mean and in fact, has a number of features that further suggest it may be interpreted as being related to some underlying ontological model. However, the ontology appears bizarre since the weak values are complex and unbounded. Here, we study weak values in the context of a recent quantum optical experiment involving two-photon interactions. The results of the experiment are reinterpreted within a `stochastic optics' model of light. The model is based on standard (Maxwell) electromagnetic theory, supplemented by stochastic fluctuations of the electromagnetic fields. We show that the conditional means of the electric field intensities correspond to the experimentally observed weak values. This is a provocative result, as it suggests that at least within this experiment, the weak value gives us information about the average of an ontological quantity (the intensity).  We study the breakdown of the stochastic optics model, which occurs outside the experimentally probed regime, and in particular in the limit where the weak value predicts `anomalous' results. 

\end{abstract}

\maketitle

\section{Introduction}

Quantum theory does not have a clear ontology that associates observables with an objective, underlying reality. Formally, we cannot make definite statements about observable quantities such as the number of photons passing through an arm of an interferometer or similarly the intensity of the electric field in that arm. Instead, we can only make predictions about the possible results of measurements that inherently disturb the system (or at least its quantum state).  Ontological theories are difficult to construct in the general case, so a good starting point for considering possible ontologies are  models that apply in particular cases, but may not (or explicitly do not) apply in others. Here, we apply such a model to a recent experiment and show that the experimentally measured values correspond to the average of the ontological entities (the electromagnetic field intensities) in the model.    

Ontological models for quantum theory should replicate predictions that defy our \emph{classical} intuition (e.g. the photoelectric effect, tunneling, contexuality and Bell-inequality violations). However,  the term \emph{classical} is not well defined and one is usually expected to understand from context what we mean by `classical intuition.'  Moreover, many phenomena that have a simple explanation within quantum theory  can be explained using a complicated, but consistent theory that one may call `classical.'

For the electromagnetic field, one may call any prediction that can be explained by Maxwell's equations `classical.' Conversely, one may call any phenomenon that cannot be explained without quantizing the electromagnetic field `quantum.' Clearly, classical electromagnetism (EM) does not require  the concept of photons, while quantum electrodynamics (QED) does. As it turns out, many of the physical effects that we associate with photons can be explained through Maxwell's equations and are, in that sense, completely classical. The most famous example is the photoelectric effect, which can be explained without photons, as long as the atoms are treated quantum mechanically \cite{Lamb1995}.

It is not surprising that there are intermediate scenarios where a particular phenomenon cannot be explained easily by Maxwell's equations, yet can be explained by a similar wave theory of light that does not include photons. Jaynes was one of the first to investigate how far one could go without quantizing the field \cite{Jaynes1973}. He was able to successfully reproduce many of the predictions of QED (such as spontaneous decay), with notable exceptions such as some details of the Lamb shift and most importantly Bell inequality violations.  In the same vein, work in ``stochastic optics" (as well as under other names such as ``stochastic electrodynamics" or ``fluctuating vacuum") starts with classical electromagnetic theory, supplemented by the hypothesis that the universe is permeated by fluctuating fields. These fluctuations, designed to replace vacuum fluctuations in quantum theory, play the role of a classical hidden variable \cite{Santos2015,Wharton2018,Penadela2005,Marshall1988,Santos1989}.

In a recent experiment  \cite{Hallaji2017} (involving two of the present authors: J.S. and A.S.)  Hallaji et al. demonstrated that if the number of photons going through one arm of an almost balanced Mach Zehnder (MZ) interferometer was measured with low precision (and low back-action) and only the subset of data corresponding to detection of a photon in the nearly dark port was analyzed, the photon-number measurement result could be surprisingly large, even suggesting that a single photon had `acted like eight photons.'  In this scenario, the number of photons measured to be in one arm of the interferometer is the \emph{weak value} \cite{Aharonov1988}, and the increase in the number of photons in that arm is an example of a phenomenon called \emph{weak-value amplification} (WVA). The counter-intuitive predictions of WVA in fact figured in the very first paper proposing weak measurements \cite{Aharonov1988}, and their potential application to precision measurement has been hotly debated in recent years \cite{Ferrie2014,sinclair2017}. Meanwhile, the physical meaning of weak-measurement results when they violate our traditional expectations remain controversial \cite{Leggett1989,Peres1989,FC100,Duck1989,Aharonov1989,Bcom,Piacentini2017a,Vaidman2017a}. Part of this debate has been centered around the fact that most WVA experiments are done with classical light and a classical coupling between two degrees of freedom on the same photon (e.g. position and polarization). The experiment of Hallaji et al. was the first photonic WVA experiment to involve a non-demolition measurement of the photon number using a second beam as a probe, and is therefore of particular significance.

Here, we show that the results of this experiment match the predictions of a stochastic optics model that takes vacuum fluctuations to be ontologically real fluctuations in electromagnetic fields.  These random fluctuations may be described by variables that remain hidden (i.e. they are not observed directly), but quantum statistics are only respected if the size of the fluctuations are chosen properly. If we take this model seriously, the weak values observed by Hallaji et al. correspond precisely to the conditional mean of the ontological field intensities.  It is instructive to consider a situation where “classical” noise greatly outweighs quantum fluctuations --  in this scenario, the astute physicist would use Bayes's rule to compute the average value of the intensity, which they would rightly interpret to be a valid estimate of the ``real state of affairs." Our result shows that in this situation, the weak value provides accurate information about the true nature of the intensity.  We are left with the interesting question of whether or not one should make a similar connection between weak values and the ontological elements `deeper' in the quantum regime.

We do not claim that our model extends beyond this experiment, and in fact we discuss regimes where the model makes predictions inconsistent with quantum theory. In particular, we find that the model fails to reproduce the predictions of quantum theory  when the fluctuations have a significant influence on the  probability of detecting a photon in the dark port (i.e. the fluctuations are large compared to the imbalance in the interferometer).  Interestingly, this limit is similar to the limit in which weak value amplification of a single post-selected photon (the experiment of ref. \cite{Hallaji2017}) would produce `anomalous results' (i.e. weak values that fall outside the range of eigenvalues\footnote{The quantum nature of anomalous weak values has been a subject of recent debate \cite{FC100,Vaidman2017,Karanjai2015}. It has been noted that there are no anomalous weak values in a deterministic classical theory \cite{Bcom} and that  there is a  direct connection between anomalous weak values and contextuality \cite{Pusey2014}.}), as we discuss in Sec. \ref{sec:discussion}.  Our work thus complements other observations of the connection between weak values and a realist model, such as \cite{Karanjai2015}, which is restricted to Gaussian quantum mechanics and thus cannot predict anomalous weak values; and \cite{Wharton2018}, which is constructed within the framework of a retrocausual model.


The outline of this paper is as follows: In Sec. \ref{sec:WV}, we discuss the formalism of weak measurement. In Sec.  \ref{sec:exp}, we describe the original proposal of Feizpour et al. \cite{Feizpour2011} and its experimental implementation performed by Hallaji et al.  \cite{Hallaji2017}.  In Sec. \ref{sec:neoclassical}, we introduce the stochastic optics model and show how it can reproduce the weak values measured in the experiment. In Sec. \ref{sec:discussion},  we discuss our results and their limitations.

 \section{Weak measurements and weak values} \label{sec:WV}
 
 One prediction of quantum theory is that if we know the state $\ket{\psi(t_0)}$  (which is to say, everything that it is possible to know according to the conventional interpretation), then for any time  $t_m\ge t_0$ there will be some measurements whose results cannot be determined {\em a priori}. This is in contrast to conventional classical theories, where indeterminism arises from incomplete knowledge \cite{Naturwissenschaften.23.844,Trimmer1980}. Moreover, in quantum theory,  knowledge of the state of the system at $t_f> t_m>t_0$ in addition to knowledge about the state at time $t_0$ allows us to make more precise statements about the results of an intermediate measurement at  $t_m$. A more complete description of the system (or at least the probability distribution for measurement results) at $t_m$ is therefore given by a two-state vector, that takes into account the states at $t_0$ and $t_f$ \cite{Aharonov1964}. We assume that the state of the system at $t_f$ was determined by an ideal measurement (i.e. a rank-1 projective measurement). We call the state associated with the result of this measurement the post-selected state.  The two-state vector for a system prepared in the state $\ket{\psi(t_0)}$ and post-selected in the state $\ket{\phi(t_f)}$ is  denoted  $\bra{\phi(t)}\ket{\psi(t)}$, where we assume that the evolution of $\ket{\psi(t)}$ forward in time and $\bra{\phi(t)}$ backwards in time is given by the Schr\"odinger equation. 
 
If we assume that the time it takes to make an intermediate measurement (at $t_m$) is very short (compared to the natural evolution of the system), we can use the two-state vector $\bra{\phi(t_m)}\ket{\psi(t_m)}$ as our  description of the system at the time of measurement.  Generally, a measurement at time $t_m$ will entangle the system with some environment so that the two-state description is no longer valid at times either before or after $t_m$.  Aharonov, Albert, and Vaidman noted, however,  that when  the measurement at $t_m$ is sufficiently weak, the two-state description can remain valid at all times \cite{Aharonov1988}. However, the result\footnote{We note that the term `result' here is somewhat controversial since in practice it is an average over many trials of an identical experiment.} of such a `weak measurement' is not the expectation value of the relevant observable, $A$. Instead, it is  an unbounded complex number that depends on  $A$ as well as the past and future boundary conditions. This number is called the weak value of $A$, 
  \begin{equation}
 \{A\}_w=\frac{\bra{\phi(t_m)}A\ket{\psi(t_m)}}{\braket{\phi(t_m)|\psi(t_m)}}.
 \end{equation}
 
 Since the measurement must be weak, a good estimate of the weak value can only be determined after many repetitions of the same experiment, with the same past and future boundary conditions. One might  think of the weak value (or similarly the two-state vector) as a variable that is hidden at time $t_0$ but is revealed at $t_f$.  If one takes this picture seriously, one might be tempted to draw an unusual, but consistent ontological picture where each observable has a weak value at all times. Such an ontology  may be  particularly satisfying since the weak value can be measured for many\footnote{Note that one can only make a finite number of  weak measurements on a single system before back-action starts to dominate. Moreover, it is still not clear if all weak values can be measured directly \cite{BCnonlocal}.}   observables simultaneously, even when the corresponding operators do not commute. Thus, we get a nearly classical picture with some unusual aspects that often appear in the form of anomalous weak values \cite{Pusey2014,Steinberg,Denkmayr2014,Dixon,Yokota2009,Lundeen2009}. Before delving into our stochastic optics model, we review the original proposal and experiment that our model describes.

 \section{Description of the experiment}\label{sec:exp}

\begin{figure}[h]
        \centering
        \resizebox{10cm}{!}{
                \includegraphics[width=\textwidth]{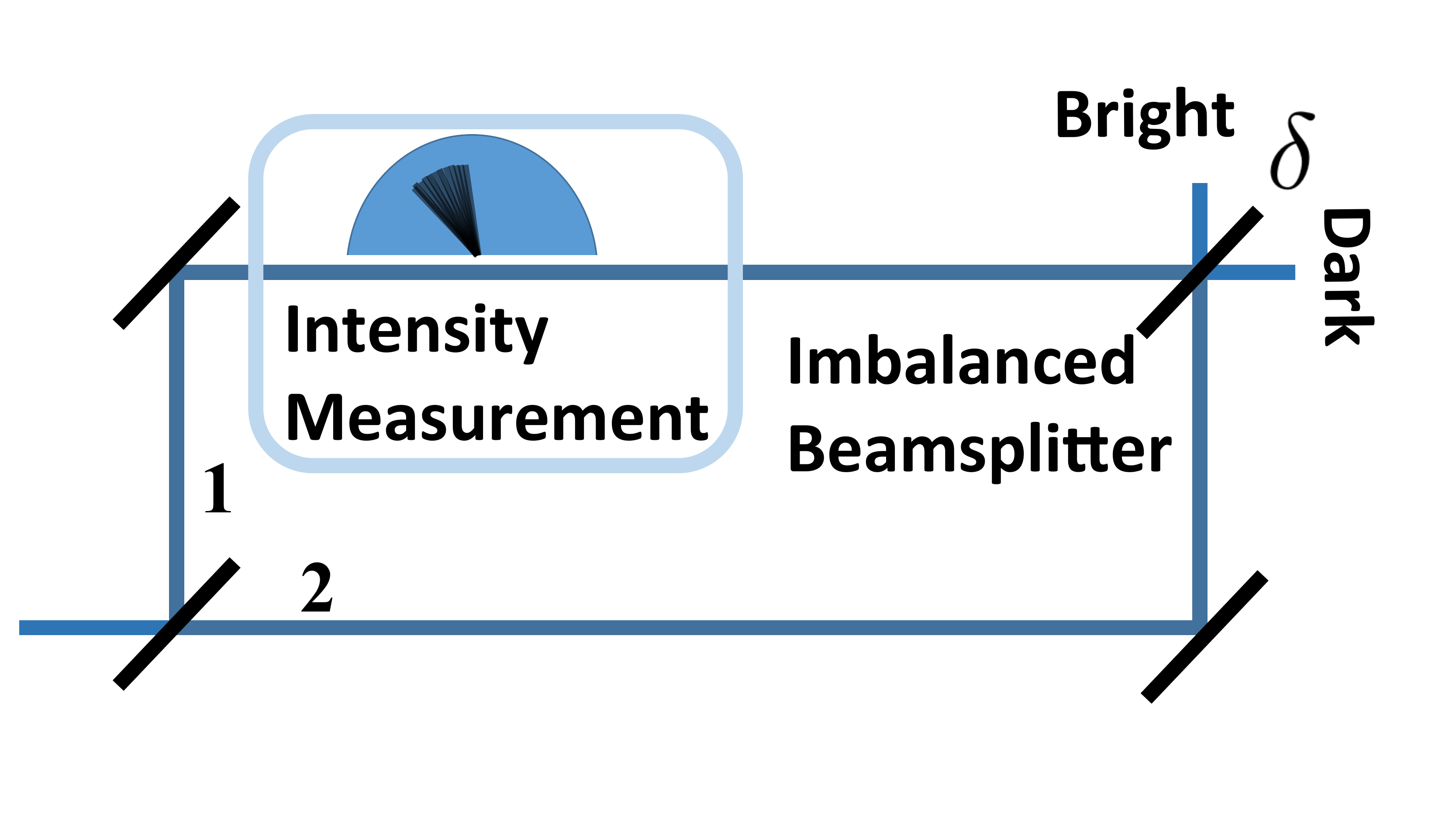}}
                \caption{Schematic of the experimental setup. The first beamsplitter is balanced, while the second is slightly imbalanced, so that a small portion, $\delta^2$, of the incoming  intensity will reach the dark port. The weak intensity measurement on arm 1 will give a mean outcome proportional to the photon number.} \label{fig1}
\end{figure}

The observations reported by Hallaji et al. \cite{Hallaji2017} constitute an experimental implementation of an earlier proposal by Feizpour et al. \cite{Feizpour2011} to amplify a single-photon non-linearity using weak measurement. In the  proposal, a single-photon Fock state passes through an imbalanced MZ interferometer and is occasionally detected at the nearly dark output port (see Fig. \ref{fig1}). Nestled in one arm (arm 1)  is a device which realizes a  non-demolition measurement of photon number (`non-demolition'  means that the photons are not absorbed, as they are with standard photon detectors). Inside the interferometer, the state can be written in the Fock basis as
\begin{equation}
\ket{i} = \frac{1}{ \sqrt{2}} \ket{1}_1 \ket{0}_2 + \frac{1}{\sqrt{2}} \ket{0}_1 \ket{1}_2,
\end{equation} 
where the subscripts $1$ and $2$ refer to arm 1 and arm 2 of the interferometer. Feizpour et al. recognized that retaining the measurement of photon number in cases where the photon was detected in the nearly dark output port and discarding the measurement results in all other cases amounted to a measurement of the weak value of photon number in one arm of the interferometer. When the final beamsplitter in Fig. \ref{fig1} is set up to have reflectivity $r$ and transmissivity $t$ defined in terms of a small parameter  $\sqrt{2} \delta = t-r$, to first order in $\delta$, the post-selected state is 
\begin{equation}
\ket{f} \approx \frac{1 + \delta}{\sqrt{2}}  \ket{1}_1 \ket{0}_2 - \frac{1-\delta} { \sqrt{2}} \ket{0}_1 \ket{1}_2,
\end{equation}
where $t \approx \frac{1+\delta}{\sqrt{2}}$ and $r \approx \frac{1-\delta}{\sqrt{2}}$ for $\delta \ll 1$. The weak value for photon number in arm $1$ of the interferometer is then   
\begin{equation}\label{eq:fockweak}
\braket{\hat{n}_1}_{w} = \frac{\braket{f|\hat{n}_1|i}} {\braket{f|i}}   \approx \frac{1}{2} + \frac{1} {2 \delta}.
\end{equation}
This value can be larger than 1, which would seem to suggest that there were on average more photons in one arm of the interferometer than there were in both arms combined. Equivalently, this result implies that there were on average a negative number of photons in the other (unmeasured) arm of the interferometer since $\braket{n_1}_{w}+\braket{n_2}_{w}=1$. This is a striking prediction, but experimental confirmation required a bright, narrow-band single-photon source. This was a significant obstacle \cite{Xing2013,Wolfgramm2008} which was overcome by substituting a weak coherent state for the single-photon Fock state. 

In their experimental implementation, Hallaji. et al. used coherent states $\ket{\alpha}$ with a  mean photon number between $|\alpha|^2=10$ and $|\alpha|^2=95$. Taking detector and other inefficiencies into account, the probability of detection at the dark port was fairly small. When this is the case, a detector firing at the dark port signals the presence of an extra photon \cite{Hallaji2016}. Therefore, detection leads to an effective post-selection of the initial state (to first order in $\delta$) with an extra photon in the dark port,
\begin{equation}
\ket{\tilde f}= \hat{a}_D^{\dagger} \ket{i} \approx \ket{\alpha}_B \hat{a}_D^\dagger\ket{\alpha\delta}_D,
\end{equation}
where the tilde is used to remind us that the state is not normalized. To calculate the weak value of photon number in arm $1$, we evaluate
\begin{equation}
\braket{\tilde f|\hat{n}_1|i} \approx \bra{\alpha}_B \bra{\delta\alpha}_D [\hat{a}_D\hat{a}_1^{\dagger} \hat{a}_1 ]\ket{\alpha}_B \ket{\alpha\delta}_D.
\label{weakvalue}
\end{equation}
The mode operators $a_1$ and $a_2$ are related to $a_B$ and $a_D$ via the beam splitter transformation
\begin{equation}
  \left({\begin{array}{c}
   \hat{a}_2\\
   \hat{a}_1  \\
  \end{array} } \right) =
  \left( {\begin{array}{cc}
   t & -r\\
  r & t \\
  \end{array} } \right) \left({\begin{array}{c}
   \hat{a}_B\\
   \hat{a}_D \\
  \end{array} } \right).
\end{equation}
Substituting into \ref{weakvalue} and retaining terms up to first order in $\delta$, we get 

\begin{equation}
\braket{\tilde f|\hat{n}_1|i}  \approx {\alpha \delta |\alpha|^2 + \alpha + \alpha \delta   \over 2 }
\end{equation}
and the weak value for photon number in arm $1$ is therefore
\begin{equation}
\label{eq:arm1Amp}
\braket{\hat{n}_1}_w \approx  {\alpha \delta |\alpha|^2 + \alpha + \alpha \delta  \over 2 \delta \alpha} ={ |\alpha|^2 \over 2} + {1 \over 2} + {1 \over 2 \delta}.
\end{equation}
Meanwhile, the mean number of photons in arm 1 without post-selection is $\braket{\hat{n}_1}=\bra{i}\hat{n}_1\ket{i}=\frac{|\alpha|^2}{2}$ so that the difference between the weak value and the mean photon number is
\begin{equation}\label{eq:diffQM}
D_I(\delta)=\braket{\hat{n}_1}_w-\braket{\hat{n}_1} \approx \frac{1}{2}+\frac{1}{2\delta}.
\end{equation}
As before, the weak value can be much larger than the mean number of photons in the initial state, implying an anomalous result for the photon number in arm 2 of the interferometer\footnote{This anomalous result can also be measured directly, simply by changing the relative phase between reflection and transmission on the final beamsplitter.}. 

\begin{figure}[h]
 \centering
\includegraphics[width=0.8\textwidth]{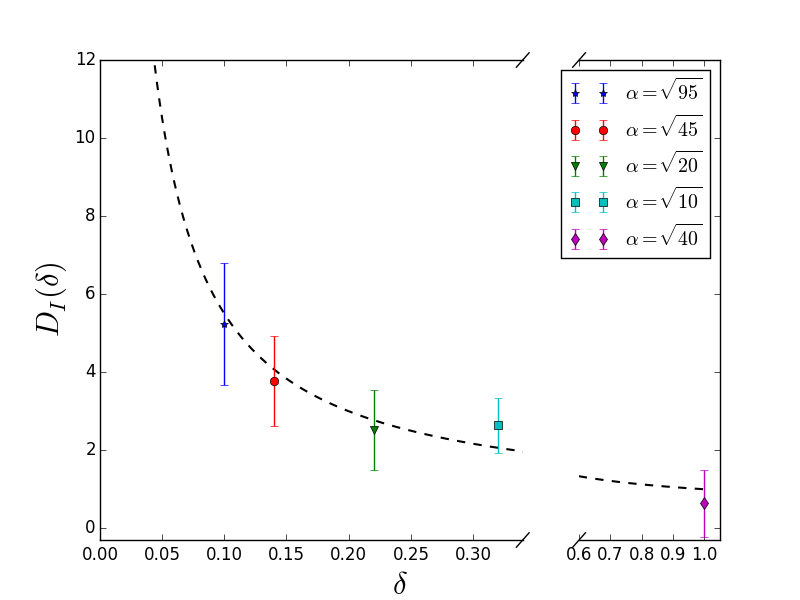}

                \caption{The results of Hallaji et al. compared with theory (dashed curve), Eq. \ref{eq:diffQM}. Each data point was taken with a different intensity for the initial state. Weak-value amplification of photon number was observed for all data points except $\delta=1$. Experimental data taken from  \cite{Hallaji2016}.
                \label{fig2}}
                
\end{figure}

 In Hallaji's experiment (see \cite{Hallaji2017} for details), the probability of the detector in the dark port firing was kept at about 25\% for all measurements. Given the collection and detection efficiency of approximately 30\%, this constrained the product $\delta^2|\alpha|^2$ (the average photon number in the dark port as $\delta$ is small) to be about 1 and as such $|\alpha|^2$ was varied between $10$ and $95$, while $\delta^2$ was varied from $0.01$ to $0.1$. An additional point at $\delta= 1$ was taken, with the beam attenuated to keep the detection rate low. The weak photon-counting measurement was implemented using a nonlinear (Kerr) interaction, which wrote a phase on a ``probe'' beam proportional to the photon number in the ``signal" pulse inside the interferometer \cite{Dmochowski2015}. The quantum uncertainty in the measurement (arising due to the  phase uncertainty of the probe beam) was around 1000 photons, which was much larger than the uncertainty due to the vacuum fluctuations in the signal. In practice, the uncertainty was even larger due to experimental imperfections.  To achieve good precision in the estimate of the weak value of photon number in arm 1 for each value of $\delta$, millions of measurements had to be taken. Hallaji reports 5 measurements of the weak value of photon number in arm 1 for different values of $\delta$. These measurements are reproduced in Fig. \ref{fig2}.


 \section{The stochastic optics model}\label{sec:neoclassical}

Our aim is to show that a simple stochastic optics model can be used to explain the results of the experiment, and moreover to study the relation between the ontological field in this model and the weak value.  Towards this end, we label the electric fields in the stochastic optics model as $E$, distinguishing them from the amplitude, $\alpha$, of the coherent quantum field.  We assume that the fields are monochromatic and choose to work in a system of units such that $|E|^2$ is a photon number (i.e. $|\alpha|^2=|E|^2$).  We extend the classical theory by assuming that the electric fields fluctuate stochastically. That is, we consider a real field whose amplitude  fluctuates around some mean value. The measurements we are discussing are of a single mode, so we treat the fluctuations as constant during each experimental run, but uncorrelated across experimental runs.  As we show below, by post-selecting on detection at the dark port we select a subset of cases where the field was in fact larger than the average. Consequently, the post-selected mean amplitude of the field tends to be much higher than the amplitude without post-selection. Note that our model does not include a measurement apparatus and as a result, its predictions are independent of any sort of measurement back-action. 

After describing the model in detail in Sec. \ref{sec:Setup} and \ref{sec:Approx}, we  show that we can reproduce the quantum predictions, Sec. \ref{Exppar}, and experimental results, Sec. \ref{sec:discussion}, by equating $E$ and $\alpha$ and matching the variance in the field fluctuations to that of a quantum coherent state (i.e. vacuum fluctuations).

\subsection{Setup}
\label{sec:Setup}
 In a classical description of the interferometer in Fig. \ref{fig1}, a field with amplitude $E_0$ is split by a balanced beamsplitter (BS) into two arms with field amplitude $E_1=E_2=\frac{E_0} {\sqrt{2}}$ and recombined on the second nearly-balanced BS, whose imbalance is characterized by the {small} parameter $\delta<<1$.  Accordingly, the field amplitude seen at the dark port is
$t E_1-r E_2=
E_0 \delta$, 
where $\delta$, $r$, and $t$ are related by $\sqrt{2} \delta = t-r$ as in Sec. \ref{sec:exp}. 
 
In the stochastic optics model, we allow $E_1$ to  fluctuate stochastically around the mean value according to a Gaussian probability distribution with RMS width $\sigma$.  For simplicity, we do not consider  fluctuations in arm $2$ of the interferometer, or in the input mode, as this does not change our main results, so $E_2 = \braket{E_2}$ and $E_0 = \braket{E_0}$. The field amplitude in arm 1 is described by the probability distribution

\begin{align}
P(E_1)=\frac{1}{\sigma\sqrt{2\pi}}e^{-\frac{(E_1-\braket{E_1})^2}{2\sigma^2}} \label{prior}.
\end{align}
A square-law detector placed at the dark port will fire with a probability that increases with the intensity, or the square of the field:
\begin{equation}P(click|E_1)\approx \eta |t E_1-r E_0/\sqrt{2}|^2,
\end{equation}
where $\eta$ is the efficiency of the detector. The full expression for $P(click|E_1)$ of course saturates at 1, but when $P(click|E_1) \ll 1$, this approximation is adequate \cite{Glauber1963}. We note that for a given distribution for $E_1$, $P(click|E_1) \ll 1$ can be ensured by keeping $\eta$ small (as was done in \cite{Hallaji2017} as described at the end of Sec. \ref{sec:exp}).  For simplicity, we omit $\eta$ from subsequent equations.
 
  We are interested in the mean of $E_1$, conditioned on the square-law detector firing. According to Bayes's rule, the probability of having field $E_1$ in arm 1 of the interferometer, given that a click has been recorded in the dark port, is $P(E_1|click)=P(click|E_1)P(E_1)/P(click)$.  Hence, the normalized probability density for $E_1$, given the dark port post-selection, is
\begin{align}
P(E_1|click)=\frac{(t E_1-r E_0 / \sqrt{2})^2}{\sigma\sqrt{2\pi}(E_0^2\delta^2+\sigma^2 t^2)}e^{-\frac{(E1-\braket{E_1})^2}{2\sigma^2}} \label{FullDist}.
\end{align}
 The mean of this new conditional probability distribution (called the posterior distribution) is the mean of the electric field amplitude $E_1$ in the cases where the detector fires. 
 
 We can already see that when the electric field distribution is shifted far from the minimum of the quadratic, the posterior distribution will be well approximated by a Gaussian multiplied by a linear slope. In this case, the posterior distribution is a Gaussian with a shifted  mean (see Fig. \ref{fig:shift}), reminiscent of the amplification that occurred in the previous section. 

\subsection{Approximation: Post-selection is dominated by imbalance in the interferometer}
\label{sec:Approx}

\begin{figure}[h]
\centering
\includegraphics[width=0.8\textwidth]{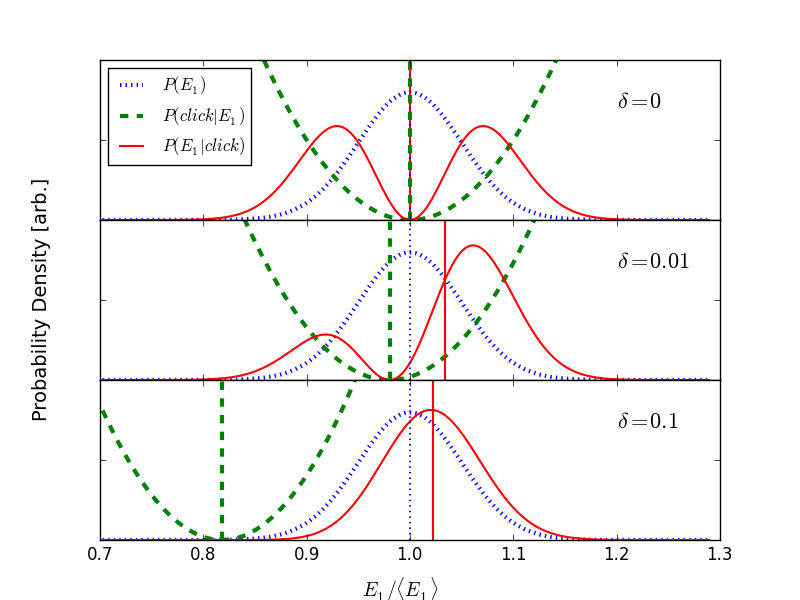}

                \caption{Illustration of the  probability distribution functions before and after post-selection:  $P(E_1)$, the  prior for $E_1$ is plotted in blue (thin dashed), the conditional probability for a click, $P(click|E_1)$, is plotted in green (thick dashed), and the shifted, posterior distribution,  $P(E_1|click)$, is plotted in red (solid). The vertical lines correspond to  the mean of the prior distribution (blue thin dashed) at $\braket{E_1}$, the mean of the posterior distribution (red solid line) given by Eq. \ref{noapprox} and the point $P(E_1|click)=0$ (green thick dashed) given by $\approx\frac{\braket{E_1}(1-\delta)}{1+\delta}$. For each figure, $\sigma=1/2$, the average field is $\braket{E_1}=10$, and $\delta$ increases from top to bottom. As $\delta$ increases, we can see that $P(E_1|click)$ better approximates a shifted Gaussian.\label{fig:shift}}
                
\end{figure}

Within our model, clicks at the dark port can occur \emph{either} due to the imbalance of the final BS or due to the fluctuations of the electric field.  We will consider the limit where the imbalance of the BS is the primary cause of the observed detections, i.e. $\sigma \ll E_0 \delta$. This is equivalent to requiring that $P(E_1)$ be narrow with respect to its deviation from the minimum of $P(click|E_1)$. Under this assumption and remembering that $t \sim \mathcal{O}(1)$, Eq. \ref{FullDist} may be approximated as
\begin{align}
P(E_1|click) \approx &\frac{(t E_1-t\braket{E_1}+E_0\delta)^2}{\sigma\sqrt{2\pi}E_0^2\delta^2}e^{-\frac{(E_1-\braket{E_1})^2}{2\sigma^2}} \label{1stApprox}
\\\approx &\frac{1}{\sigma\sqrt{2\pi}}\bigg{(}1+\frac{2t (E_1-\braket{E_1})}{E_0\delta}+\frac{t^2 (E_1-\braket{E_1})^2}{E_0^2\delta^2}\bigg{)}e^{-\frac{(E_1-\braket{E_1})^2}{2\sigma^2}} \label{expanded},
\end{align}
where in the first line we have used the fact that $r=t-\sqrt{2}\delta$ and in the second line merely simplified terms.    The second and third terms will only have significant contributions to $P(E_1|click)$ when  $ |E_1-\braket{E_1}|\lesssim\sigma$.  Hence, disregarding the third term amounts to the same approximation as in Eq. \ref{1stApprox}.  Furthermore, the second term is again small compared to unity, and as such, using the fact that $e^a\approx1+a$ for $a<<1$, we can further simply,
\begin{align}
P(E_1|click)\approx\frac{1}{\sigma\sqrt{2\pi}}exp{\left[-\frac{(E_1-\braket{E_1})^2}{2\sigma^2}+\frac{2t (E_1-\braket{E_1})}{E_0\delta} \right]}.
\end{align}
Combining terms and completing the square in the exponential, we find that 
\begin{align}
P(E_1|click)\approx\frac{1}{\sigma\sqrt{2\pi}}exp{\left[-\frac{2t^2\sigma^2}{E_0^2\delta^2}\right]}exp{\left[-\big{(}E_1-\braket{E_1}-2t\sigma^2/E_0\delta\big{)}^2
\big{/}2\sigma^2\right]} \label{ShiftedGauss}.
\end{align}
As promised, Eq. \ref{ShiftedGauss} describes a shifted Gaussian as compared to the prior distribution in 
Eq. \ref{prior}, as illustrated in Fig. \ref{fig:shift}. The shift in the mean scales like $1/\delta$ in agreement with the amplified weak value predicted by quantum theory.

In the experiment, detectors measure field intensities. The intensity in arm 1 should be proportional to $|E_1|^2$ and the difference in the intensities with and without post-selection should be proportional to\footnote{Allowing the field in arm 2 to fluctuate with variance $\sigma_2$ will modify Eq. \eqref{noapprox} to 
\begin{equation*} 
D_I(\delta)=\left[\frac{4\sigma^2 t}{\sqrt{2}\delta}+\frac{2t^2\sigma^4}{E_0^2\delta^2}
\right]\bigg/\left[ 1+\frac{t^2\sigma^2+r^2\sigma_2^2}{E_0^2\delta^2}
\right].
\end{equation*}
In our approximation this yields  the same result when $\sigma_2=\sigma$.}  
\begin{align}
D_I(\delta)&=\int{ P(E_1|click)|E_1|^2dE_1}-\int{P(E_1)|E_1|^2dE_1}\\
&=\left[\frac{4 \sigma^2t}{\sqrt{2}\delta}+\frac{2t^2\sigma^4}{E_0^2\delta^2}
\right]\bigg/\left[ 1+\frac{t^2\sigma^2}{E_0^2\delta^2}
\right]\label{noapprox},
\end{align}
which is derived by integrating directly using Eqs. \ref{prior} and \ref{FullDist}. Note that this can be rewritten, 
\begin{equation}
\label{eq:shiftedmean}
    D_I(\delta) = {\frac{4 \sigma^2 t }{\sqrt{2} \delta}}{\left[\frac{1+B}{1+A} \right] }\approx{\frac{4 \sigma^2 t }{\sqrt{2} \delta}}(1 - A + B) \approx\frac{2 \sigma^2(1+\delta)}{\delta},
\end{equation}
where $A \approx \frac{\sigma^2 }{2 E_0^2 \delta^2}$ and $B \approx \delta A$ (to lowest order in $\delta$). In the approximation, we have made the simplification that $t\approx(1+\delta)/\sqrt{2}$ ($r\approx(1-\delta)/\sqrt{2}$), as $\delta\ll 1$, and that $A \ll 1$, which corresponds to the approximation, $\sigma \ll E_0 \delta$, that we made when dealing with field amplitudes earlier in Sec \ref{sec:Approx}.

We are now in a position to compare our result with the quantum prediction, Eq. \ref{eq:diffQM}. Qualitatively, the stochastic optics theory shows good agreement, predicting a shift of the form $constant \times (1+1/\delta)$. In order to quantitatively compare the model to quantum theory, we will in the next section relate parameters in the stochastic optics model, ($\sigma, E_0)$, to the more familiar quantities in the quantum treatment for a coherent state of light, (${1 \over 2}, \alpha$).  

\subsection{Comparing to the experiment}
\label{Exppar}

Recall that we chose our units so that $E_0 = \alpha$ and so Eq.  \ref{eq:shiftedmean} can be compared directly with Eq. \ref{eq:diffQM}. We now choose $\sigma$ to coincide with the vacuum fluctuations in the same units, that is  $\sigma=1/2$  \cite{Gerry2005}. Thus the field has the statistics of a coherent state.
We can now evaluate Eq. \ref{eq:shiftedmean}, 
\begin{equation} \label{eq:mainresult}
D_I(\delta)\approx\frac{1}{2}\left[1+\frac{1}{\delta}\right]
\end{equation}
and arrive at the same result that Hallaji et al. found  using the weak value formalism, most importantly the amplification factor of $\frac{1}{\delta}$. {This is the main result of the stochastic optics model: under the above approximations, weak-value amplification of photon number is successfully modeled by a theory with classical electric fields and Gaussian {vacuum} fluctuations. We emphasize that in this model, the electric field is the ontological entity and the conditional mean of the intensity is given by the weak value.}

\section{Discussion}\label{sec:discussion}
{The main technical result, Eq. \ref{eq:mainresult}, shows that within our model, the conditional mean of the `real' intensity is the weak value and exhibits weak-value amplification.  This agreement between the predictions of the stochastic optics model and the predictions of quantum theory only arises after several assumptions have been made. In the next section, we examine these assumptions more carefully, identify a regime of validity for the stochastic optics model (defined as a regime where the model's predictions concur with the predictions of quantum theory), and show that the experiments performed by Hallaji et al. fall within this regime.}

\subsection{Exploring the mathematical assumptions}
\begin{figure}[h]
\centering
\begin{minipage}{0.49\textwidth}
	\centering
	\includegraphics[width=1.0\textwidth]{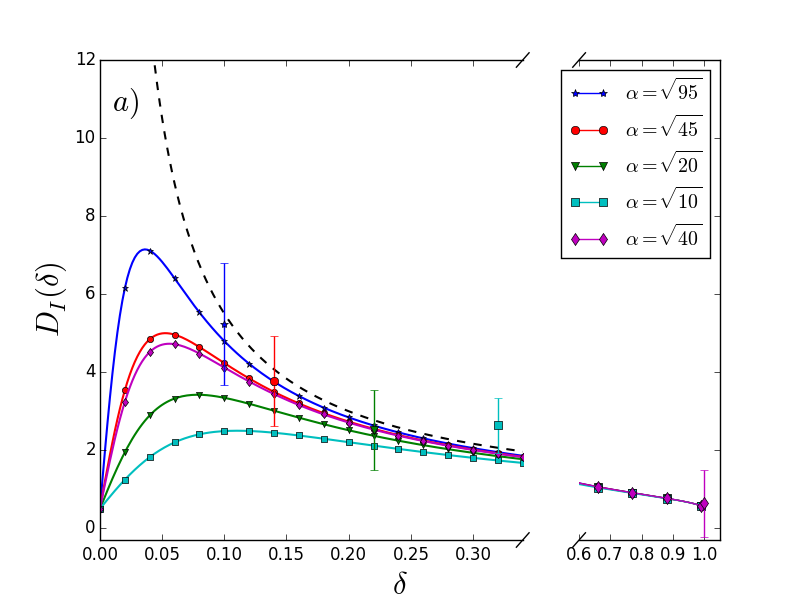}
\end{minipage}
\begin{minipage}{0.49\textwidth}
	\centering
	\includegraphics[width=1.0\textwidth]{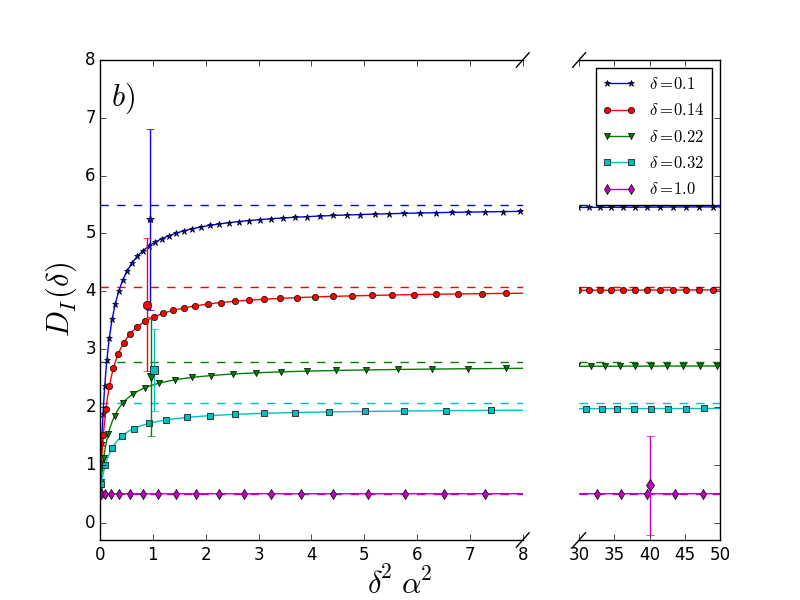}
\end{minipage}

\caption{The shift in the intensity (number of photons) due to post-selection as a function of: $a)$ $\delta$, the imbalance in the interferometer b) $\delta^2\alpha^2$, the mean intensity at the dark port. Each solid curve depicts the stochastic optics theory for $a)$ a fixed value of $\alpha$ and $b)$ a fixed value of $\delta$, while the dashed curves represent the weak values.  Finally, the markers with errors show the experimental data from \cite{Hallaji2016}, with colors (symbols) indicating the corresponding stochastic optics curve. As $\delta^2\alpha^2$ becomes large, the shift converges to the weak value. The BS imbalance  dominates the probability for post-selection when $\left[\frac{\sigma}{ \alpha\delta}\right]^2\ll1$ ($\sigma = 1/2$ in both plots). This corresponds to the scenario when the intensity at the dark port is larger than the fluctuations due to the vacuum. The $\delta=1$ (magenta diamonds) line is the base shift (a photon added in a superposition of both arms) while all values above $D_I=1/2$ are a result of weak-value amplification.}
\label{fig4}
\end{figure}

In quantum theory the conditions that lead to a prediction of weak-value amplification (Eq. \ref{eq:diffQM})  are that the post-selection be dominated by the imbalance of the interferometer rather than by the measurement back-action \cite{Aharonov1988,Salvail2013}. The stochastic optics model does not include a photon-number measurement, and therefore has no back-action. On the other hand, stochastic fluctuations can lead to clicks in the dark port even if the interferometer is balanced, unlike in quantum theory. The condition that the post-selection must be dominated by the imbalance, rather than by the fluctuations of the field amplitude is the same condition used to derive Eq. \ref{eq:shiftedmean}. Mathematically, this is expressed by
\begin{equation}
\label{eq:condition}
\left[\frac{\sigma}{ \delta E_0}\right]^2\ll1.
\end{equation}
This implies that the stochastic optics model will cease to agree with the quantum prediction of weak-value amplification if the field amplitude is too low ($E_0$ or $\alpha \rightarrow 0$). Experimental tests outside the regime of Eq. \ref{eq:condition} should yield a violation of the model's predictions. This is illustrated in Fig. \ref{fig4}, where we compare the quantum prediction for the difference in intensity, $D_I(\delta)$, (dashed lines) to the stochastic optics prediction (bold lines). As expected, for low mean photon number in the dark port, $\delta^2\alpha^2$, the stochastic optics theory starts to deviate from the quantum prediction. At  $\delta=1$ i.e. $r=-t=\frac{-1}{\sqrt{2}}$ (magenta diamonds) all the light reaches the `dark port', so that post-selection results in the addition of a single photon in an equal superposition of both arms (i.e. $D_I(1)=1/2$). Anything above this line is associated with weak-value amplification.

Hallaji et al. report on four measurements where both $\delta$ and $\alpha$ are varied and the expected amplification effect was observed, and one where it was not expected to be observed ($\delta = 1$).{ In the small $\delta$ case, the values of $\alpha$ were chosen such that $\delta\alpha\approx 1$ so that a sufficient number of photons would reach the dark port. } For the 5 reported measurements the assumption that $\left[\frac{\sigma}{ \delta\alpha}\right]^2 \ll 1$ is good,
\begin{equation}
 \left[\frac{\sigma}{ \delta\alpha}\right]^2\approx [0.16, 0.18, 0.19, 0.21,0.01],
\end{equation}
and therefore the data fall within the regime where stochastic optics and quantum predictions agree. Taking the stochastic optics model seriously, one can consistently interpret Hallaji's results as reporting on a measurement of the average of the ontological intensity. 


\subsection{Relationship between the weak value and the ontological elements of the model}
Assigning $\sigma=1/2$ was motivated by the fact that Hallaji et al. studied weak-value amplification for a pure coherent state. Alternatively, one could use states with additional ``technical noise" (this would be a state with larger than minimal uncertainty). The model predicts a similar weak-value amplification effect, with the shift in intensity given by the more general Eq. \ref{eq:shiftedmean}. There is in this slightly modified experiment, a significant interpretational difference. Any state with added technical noise is a mixed state which can be decomposed as a probabilistic mixture of states with different amplitudes, or equivalently, as a stochastically varying electric field. We therefore do not need to invoke our stochastic optics theory to introduce fluctuations in the field amplitude; they are already present in the conventional description.  In this regime, there would be no doubt as to the relationship between the conditional mean of the field intensity and the underlying ontology. As strange as the weak-value results might seem at first glance, a conventional physicist would interpret these very expressions as accurate representations of the conditional mean of the intensity upon successful post-selection, explained by the well-known phenomenon of photon ``bunching'' in thermal states. Our result shows, however, that the experimental relevance of this quantity persists even when the input is a pure (e.g. coherent) state, supporting the stochastic optics view that the ``true'' field may be thought of as a stochastic variable.  This interpretation fails only once the model itself breaks down (most relevantly when $\alpha \rightarrow 0$). 

The breakdown of the stochastic optics model is not by itself surprising. The simple, realist model proposed here cannot possibly reproduce the anomalous (negative) weak values that appear in the regime where $\alpha \rightarrow 0$. What was not obvious at the outset was that a simple model could reproduce any of the weak values in such a system. In fact, our model is able to fully capture the counter-intuitive amplification in a broad regime before the weak value becomes strictly anomalous. Whereas the weak value can be negative when   $|\alpha|^2 + 1 \leq 1/\delta$, our model breaks down when $\alpha \ll 1/(2\delta)$. The physical interpretation of this more relaxed condition is that post-selection in the dark port be dominated by the imbalance of the interferometer and not by the vacuum fluctuations of the real fields. This insight suggests a possible path forward towards the goal of constructing a (possibly contextual) hidden-variable model, which can reproduce anomalous weak values.

\section{Conclusion}
 We have shown that the weak-value amplification experiment performed by Hallaji et al. agrees with the results of a simple and intuitive hidden variable model based on classical electrodynamics, with stochastic field fluctuations in place of vacuum fluctuations.  Our model shows that the experimental results can be explained without quantizing the electromagnetic field. Furthermore, it allows us to think about the weak value without discussing measurements. In our model, weak values have a natural interpretation: they are the adjusted mean of the ontological field intensity following successful post-selection. While not a complete theory, the model does suggest that the weak value may be interpreted as a conditional mean in a wider range of scenarios than conventionally thought.
 
The significance of the weak value in the model is particularly apparent when imprecise weak measurements are compared with precise strong measurements. One might naively expect a strong measurement in the lab to be an accurate measurement of the ontological field in a stochastic optics model. Consequently,  one would expect the mean result of a post-selected strong measurement to be the weak value. Of course, this is incorrect. Quantum theory tells us that strong measurements are accompanied by measurement back-action which perturbs the state. Supplementing our model with the extra baggage required to describe strong measurements goes beyond the scope of this work. It suffices to note that within such a supplemented model strong measurements still do not reveal the state of the ontological field. A similar argument can be made for single-shot weak measurements, since the uncertainties are large. However, many repetitions can be used to get an estimate of the underlying ontological elements. As such, weak measurements are better tools than strong measurements for probing the underlying ontology. These results are consistent with other ontological models such as Bohm theory, where weak measurements can be used to reconstruct particle trajectories \cite{Mahler2016,Wiseman2007} and provide a starting point for a more general, realist interpretation of weak values.

\section{Acknowledgements}
This research was supported by NSERC, CIFAR, and the Fetzer Franklin Fund of the John E. Fetzer Memorial Trust.

\bibliographystyle{unsrt}
\bibliography{neoclassical}

\end{document}